# QUANTIFICATION DU RISQUE SYSTEMATIQUE DE MORTALITE POUR UN REGIME DE RENTES EN COURS DE SERVICE

**Frédéric PLANCHET*** **Laurent FAUCILLON** **Marc JUILLARD**

***ISFA – Laboratoire SAF***
*Université Claude Bernard – Lyon 1*
*50 avenue Tony Garnier*
*69007 LYON*
*FRANCE*

***WINTER & Associés***
*18 avenue Félix Faure*
*69007 Lyon*
*FRANCE*

RESUME

L'objectif de ce travail est de proposer un modèle réaliste et opérationnel pour quantifier la part de risque systématique de mortalité incluse dans un engagement de retraite. Le modèle présenté est construit sur la base d'un modèle de Lee-Carter. Les tables prospectives stochastiques ainsi construites permettent de projeter l'évolution des taux de décès aléatoires dans le futur et de quantifier la part de risque non mutualisable dans l'engagement d'un régime de rentes.

ABSTRACT

The aim of this paper is to propose a realistic and operational model to quantify the systematic risk of mortality included in an engagement of retirement. The model presented is built on the basis of model of Lee-Carter. The stochastic prospective tables thus built make it possible to project the evolution of the random mortality rates in the future and to quantify the systematic risk of mortality.

---

* fplanchet@winter-associes.fr

## Sommaire

# 1. INTRODUCTION

Depuis une dizaine d'années, les tables de mortalité utilisées pour le provisionnement et la tarification des rentes viagères s'attachent à prendre en compte la dérive tendancielle de la mortalité, qui se traduit par une croissance régulière de l'espérance de vie. On constate ainsi en France depuis une trentaine d'année une augmentation de l'espérance de vie à la naissance d'environ un trimestre par an. La nécessité d'utiliser des tables de mortalité prospectives intégrant ce phénomène de dérive a été prise en compte par le législateur et des tables « de génération » ont été homologuées il y a une dizaine d'années. Ces tables, obtenues sur la base de la mortalité de la population féminine sur la période 1961-1987, servent depuis le 1er juillet 1993 à la tarification et au provisionnement de contrats de rentes viagères (immédiates ou différées). Elles sont en cours d'actualisation et des nouvelles tables devraient voir le jour en 2006.

Des études récentes (CURRIE et al. [2004]) font toutefois apparaître que l'évolution du taux instantané de mortalité présente, aux différents âges, des variations erratiques autour de la tendance qui se dégage, variations non expliquées par les fluctuations d'échantillonage ; CAIRNS et al. [2004] présente une analyse détaillée de ce phénomène. Sur les données INED utilisées dans le présent travail (*cf.* la section 2.2.1 *infra* pour le descriptif de ces données reprises de MESLE et VALLIN [2002]) on constate le même phénomène, illustré par la figure 1 :

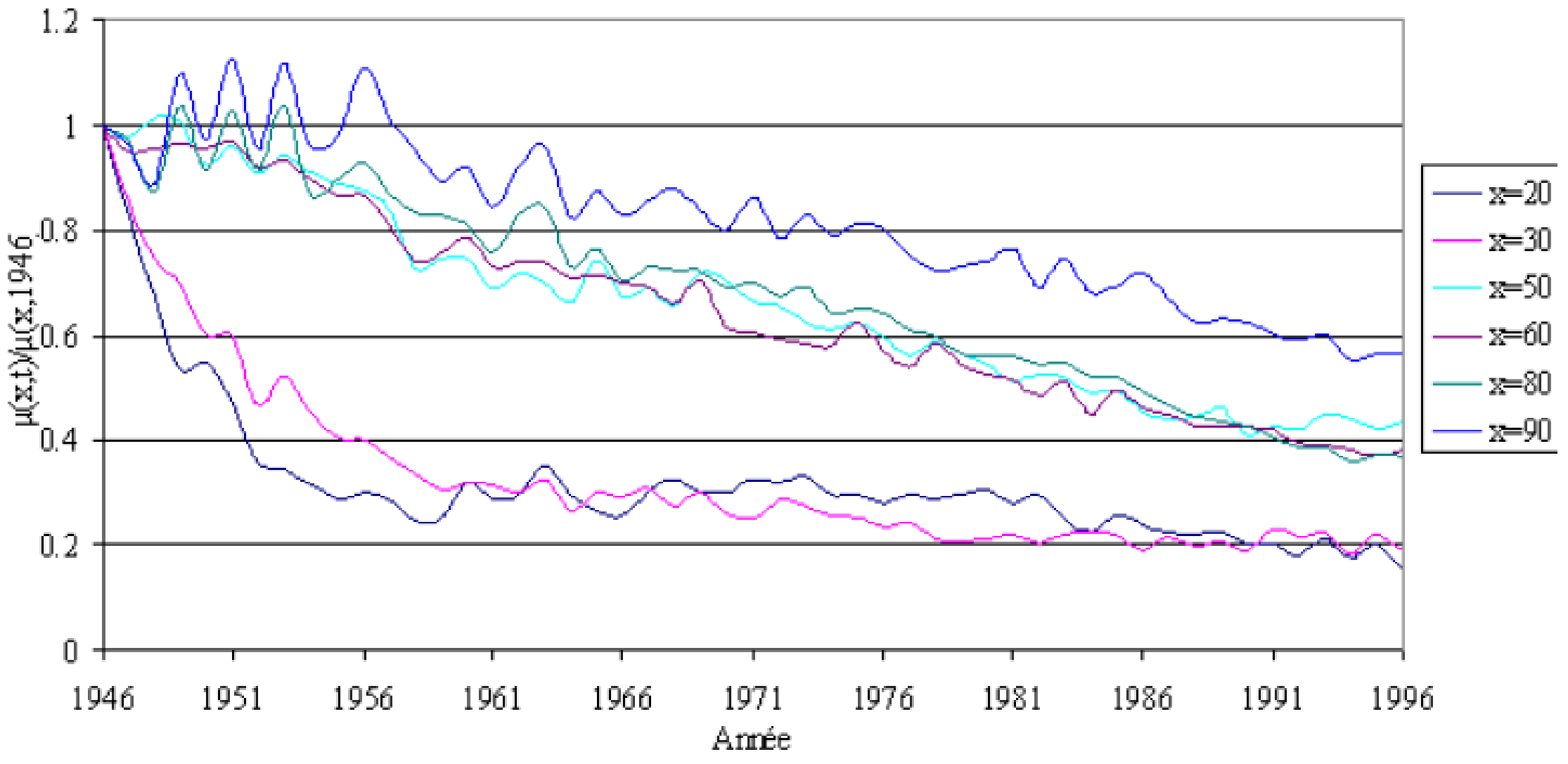

Fig. 1 : *Fluctuations des taux de décès autour de la tendance de long terme*

Ces variations affectent de manière systématique l'ensemble des individus d'âge fixé et ne se mutualisent pas ; elles font donc courir un risque potentiellement important à un régime de rentes, dont l'équilibre technique est construit sur la mutualisation du risque de survie de ses membres.

On est donc conduit à rechercher une modélisation capable de rendre compte de ces fluctuations autour de la valeur tendancielle et d'en tirer des conséquences sur le niveau des provisions mathématiques que doit constituer le régime pour assurer son équilibre technique. De manière plus précise, il importe de quantifier la part dans le risque total auquel est exposé le régime (selon une mesure à définir) de la composante non mutualisable.

Les modèles stochastiques de mortalité fournissent un outil bien adapté à cette analyse. Ils proposent en effet de considérer que le taux de mortalité futur $\mu(x,t)$ est lui-même aléatoire, et que donc $\mu(x,t)$ est un processus stochastique (comme fonction de $t$ à $x$ fixé). Le taux de mortalité effectivement observé pour un âge et une année donnés est alors la réalisation d'une variable aléatoire : on peut noter l'analogie avec les méthodes de lissage bayésiennes (sur ces méthode de lissage, on pourra consulter TAYLOR [1992]).

Dans la littérature, les modélisations stochastiques des phénomènes de mortalité sont nombreuses. On ne considère pas ici les modèles d'inspiration financière nés de la problématique de tarification de dérivés de mortalité. Le lecteur intéressé par ces approches pourra consulter DAHL [2004] et SCHRAGER [2006]. On notera simplement que l'ajustement de ces modèles à des données de mortalité n'est pas immédiat et que, pour la situation qui nous préoccupe ici, la non prise en compte des dépendances entre les âges est problématique (ces modèles proposent une dynamique d'évolution en fonction du temps pour chaque âge sans prendre en compte les corrélations existant enre les âges proches).

Plusieurs modèles classiques sont de fait des modèles stochastiques ; en premier lieu, les lissages bayésiens, et en particulier le modèle de Kimeldorf-Jones (KIMELDORF et JONES [1967]) entrent dans cette catégorie. Toutefois, ils sont élaborés essentiellement dans la perspective du lissage d'une table du moment, sans intégrer naturellement la dimension prospective essentielle dans l'analyse de l'engagement d'un régime de rentes.

Les modèles récents de construction de tables prospectives, comme le modèle de Lee-Carter (voir notamment LEE et CARTER [1992], LEE [2000], SITHOLE et al. [2000]) ou les modèles poissoniens (*cf.* BROUHNS et al. [2002] et PLANCHET et THEROND [2006] pour une présentation et une discussion de ces modèles), sont également des cas particuliers de modèles stochastiques, bien qu'ils soient à l'origine élaborés pour construire des extrapolations (temporelles) de la surface $\mu(x,t)$ déterministe. Après un ajustement des taux passés, les taux de mortalité pour les années futures se déduisent directement de l'extrapolation de la composante temporelle (paramétrique ou non paramétrique) du modèle prospectif retenu (on peut évidemment critiquer cette approche purement extrapolative ; on pourra par exemple consulter GUTTERMAN et VANDERHOOF [1999] sur ces questions).

On retient dans cette étude le modèle de Lee-Carter, qui permet aisément de produire des surfaces de mortalité stochastiques et est par ailleurs devenu un standard pour la construction de tables prospectives. La variante log-Poisson (*cf.* BROUHNS et al. [2002]) conduit à des résultats très proches.

Après avoir construit un jeu de tables prospectives sur des données nationales à l'aide de ce modèle, nous l'utilisons pour calibrer la composante aléatoire pour le risque systématique et appliquons le modèle ainsi obtenu pour déterminer le niveau de l'engagement d'un régime de rentes.

Les données utilisées pour les applications numériques sont reprises de PLANCHET et THEROND [2004].

Le présent article est inspiré d'un travail de recherche effectué dans le cadre de l'Institut de Science Financière et d'Assurances de l'Université Lyon 1 par FAUCILLON et al. [2006].

## 1.1. CARACTERISTIQUES DU PORTEFEUILLE DE RENTES

Dans la suite, nous utiliserons pour les applications numériques un portefeuille constitué de 374 rentiers de sexe féminin âgés en moyenne de 63,8 ans au 31/12/2005. La rente annuelle moyenne s'élève à 5,5 k€. La figure 2 *infra* présente les flux de prestations espérés en fonction du temps obtenu à partir de la table de mortalité TV 2000.

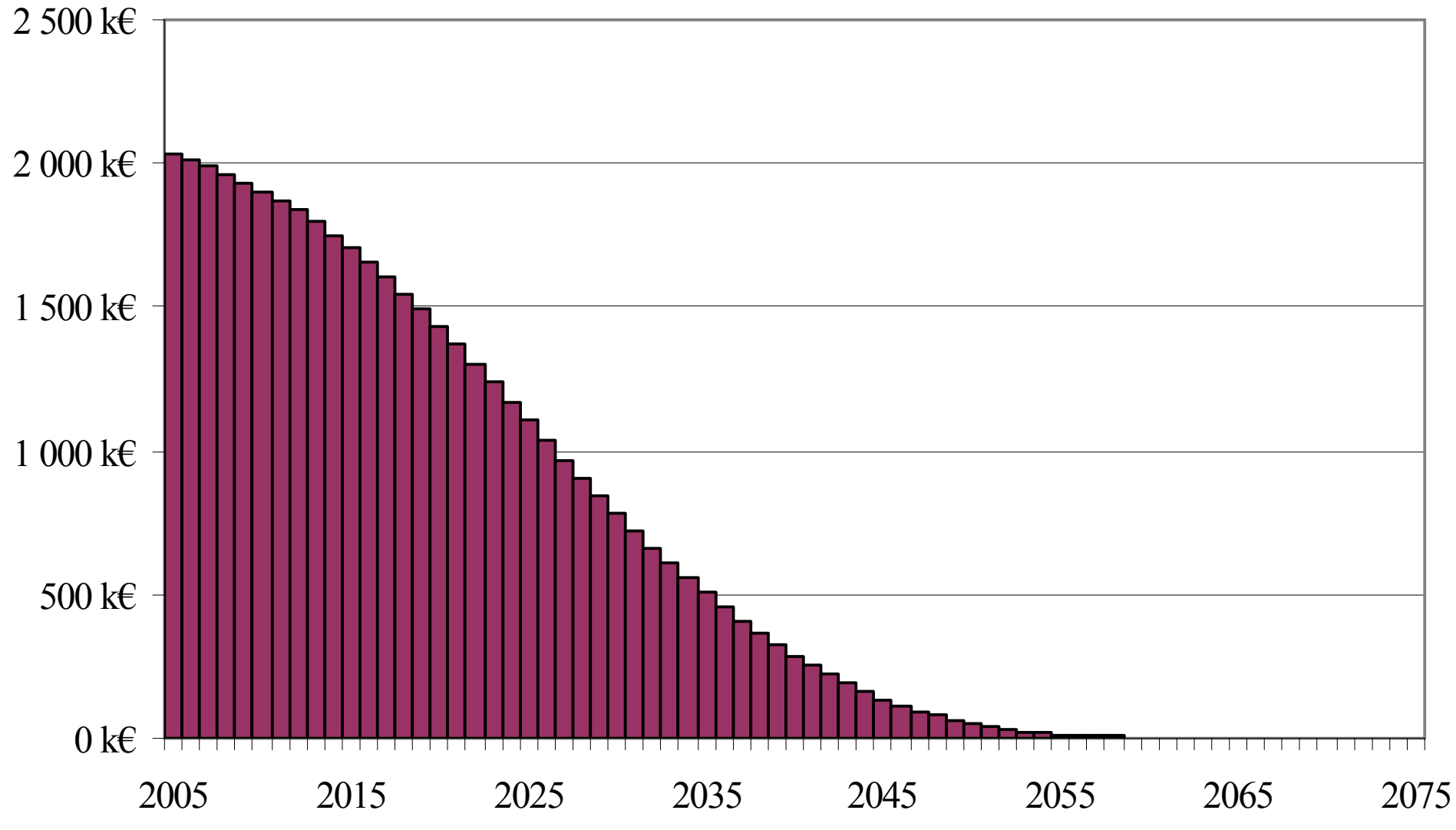

Fig. 2 : *Flux de prestations espérées*

Avec un taux d'escompte des provisions de 2,5 %, la provision mathématique initiale calculée avec la table prospective construite à la section 2.2.1, s'élève à 36,7 M€. On peut noter que dans PLANCHET et THEROND [2004], les calculs menés avec la table du moment féminine TV2000 conduisaient à une provision de 32,8 M€. Cela met incidemment en évidence l'importance d'une approche prospective de la mortalité pour déterminer le niveau de l'engagement sans le sous-estimer exagérément.

## 1.2. NOTATIONS

Nous noterons dans la suite de l'article :

- ✓ $L_0$ le montant des provisions mathématiques à la date initiale,
- ✓ $\tilde{F}_t$ le flux de prestation (aléatoire) à payer à la date *t*,
- ✓ $(\Phi_t)$ la filtration décrivant l'information disponible à la date *t*,
- ✓ *i* le taux (discret) d'escompte des provisions mathématiques,
- ✓ **J** l'ensemble des individus,
- ✓ $x(j)$ l'âge en 0 de l'individu *j* et $r_j$ le montant de sa rente annuelle.

## 1.3. PROBLEMATIQUE

Le portefeuille est exclusivement constitué de rentes en cours de service supposées non réversibles. En 0, l'assureur estime la suite de flux probables de sinistres que nous noterons $(F_t)_{t\geq 1}$ :

$$F_t = \mathbf{E}\left[\tilde{F}_t \mid \Phi_0\right], \tag{1}$$

avec :

$$\tilde{F}_t = \sum_{j\in\mathbf{J}} r_j * \mathbf{1}_{]t;\infty[}\left(T_{x(j)}\right), \tag{2}$$

où $T_{x(j)}$ désigne la date de décès (aléatoire) de la tête d'âge $x(j)$. Avec les notations classiques de l'assurance vie (*cf.* PETAUTON [1996]) :

$$F_t = \sum_{j\in\mathbf{J}} r_j * \frac{l_{x(j)+t}}{l_{x(j)}}. \tag{3}$$

A partir de cette estimation, l'assureur détermine la provision mathématique $L_0$ :

$$L_0 = \sum_{t=1}^{\infty} F_t (1+i)^{-t} . \tag{4}$$

Dans la suite on analysera le montant aléatoire de l'engagement, soit la variable aléatoire :

$$\Lambda = \sum_{t=1}^{\infty} \tilde{F}_t (1+i)^{-t} = \sum_{t=1}^{\infty} \frac{1}{(1+i)^t} \sum_{j \in \boldsymbol{J}} r_j * \boldsymbol{1}_{]t;\infty[}\left(T_{x(j)}\right) \tag{5}$$

Cette variable aléatoire est telle que $E(\Lambda) = L_0$. Pour une analyse détaillée de la loi de $\Lambda$ lorsque la mortalité est connue (avec éventuellement des taux d'intérêt stochastiques), on se reportera à MAGNIN et PLANCHET [2000].

Les propriétés de $\Lambda$ seront examinées dans un premier temps avec l'hypothèse classique que la mortalité future est connue (déterministe), via l'utilisation d'une table prospective, puis, dans un second temps, dans un contexte de mortalité stochastique. Dans ce second cas on aura donc à considérer des surfaces de mortalité stochastiques $\Pi$ et, une surface étant donnée, la loi conditionnelle de l'engagement $\Lambda | \Pi$.

## 2. CAS D'UNE MORTALITE DETERMINISTE

### 2.1. MODELE DE MORTALITE

On choisit de modéliser les taux instantanés de sortie (taux de hasard) ; le lien avec les données discrètes (annuelles) disponibles est fait sous l'hypothèse classique de constance du taux de hasard sur chaque carrée du diagramme de Lexis (*cf.* PLANCHET et THEROND [2006]), qui conduit à poser $\mu_{xt} = -\mathbf{ln}(1 - q_{xt})$.

Le modèle retenu pour constuire les tables prospectives est le modèle de Lee-Carter. Il s'agit d'une méthode d'extrapolation des tendances passées initialement utilisée sur des données américaines, qui est devenue rapidement un standard (voir l'article original LEE et CARTER [1992]). La modélisation proposée par ces auteurs pour le taux instantané de mortalité est la suivante :

$$\mathbf{ln}\, \mu_{xt} = \alpha_x + \beta_x k_t + \varepsilon_{xt} , \tag{6}$$

en supposant les variables aléatoires $\varepsilon_{xt}$ indépendantes, identiquement distribuées selon une loi $N(0,\sigma^2)$ ; l'idée du modèle est donc d'ajuster à la série (doublement indicée par *x* et *t*) des logarithmes des taux instantanés de décès une structure paramétrique (déterministe) à laquelle s'ajoute un phénomène aléatoire ; le critère d'optimisation retenu est la maximisation de la variance expliquée par le modèle, ou, de manière équivalente, la minimisation de la variance des erreurs.

Le paramètre $\alpha_x$ s'interprète comme la valeur moyenne des $\mathbf{ln}(\mu_{xt})$ au cours du temps. On vérifie par ailleurs que $\frac{d\,\mathbf{ln}(\mu_{xt})}{dt} = \beta_x \frac{dk_t}{dt}$ et on en déduit que le coefficient $\beta_x$ traduit la sensibilité du logarithme de la mortalité instantanée à l'âge *x* par rapport à l'évolution générale $k_t$, au sens où $\frac{d\,\mathbf{ln}(\mu_{xt})}{dk_t} = \beta_x$. En particulier, le modèle de Lee-Carter suppose la constance au cours du temps de cette sensibilité.

Afin de rendre le modèle identifiable, il convient d'ajouter des contraintes sur les paramètres ; on retient en général les contraintes suivantes :

$$\sum_{x=x_m}^{x_M} \beta_x = 1 \text{ et } \sum_{t=t_m}^{t_M} k_t = 0 . \tag{7}$$

On obtient alors les paramètres par un critère de moindres carrés (non linéaire) :

$$\left(\hat{\alpha}_x, \hat{\beta}_x, \hat{k}_t\right) = \mathbf{arg\,min} \sum_{x,t} \left(\mathbf{ln}\, \mu_{xt}^* - \alpha_x - \beta_x k_t\right)^2 . \tag{8}$$

Il convient donc de résoudre ce programme d'optimisation, sous les contraintes d'identifiabilité. Le nombre de paramètres à estimer est élevé, il est égal à $2 \times (x_M - x_m + 1) + t_M - t_m - 1$. Les algorithmes de résolution ne sont pas repris ici, le lecteur intéressé pourra consulter les nombreux articles décrivant ce modèle, par exemple BROUHNS et al. [2002]. Une présentation détaillée est également proposée dans PLANCHET et THEROND [2006]. Une discussion des limites de ce modèle est proposée dans LELIEUR et PLANCHET [2006] et dans SERANT [2005].

Une fois ajustée la surface de mortalité sur les données passées, il reste à modéliser la série $(k_t)$ pour extrapoler les taux futurs ; pour cela, on utilise en général un modèle ARIMA[1], mais toute autre modélisation de série temporelle peut être utilisée. Toutefois, compte tenu de l'allure obtenue sur les données disponibles :

---

[1] En suivant la démarche de Box et Jenkins.

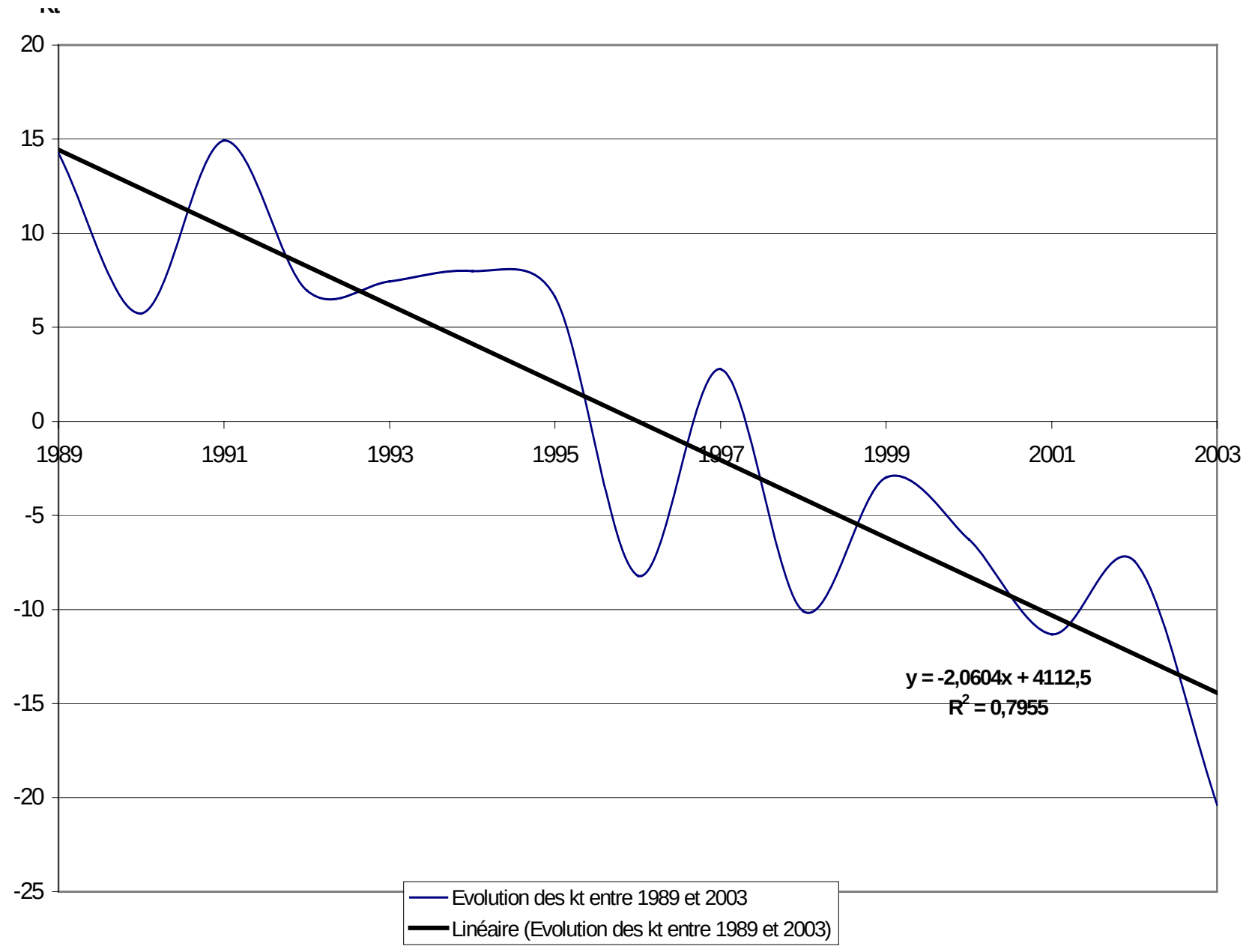

Fig. 3 : *Evolution du paramètre k(t) avec le temps*

la modélisation la plus simple que l'on puisse imaginer est une régression linéaire de ces coefficients en supposant une tendance affine :

$$k_t^* = at + b + \gamma_t, \tag{9}$$

avec $(\gamma_t)$ un bruit blanc gaussien de variance $\sigma_\gamma$. On obtient ainsi des estimateurs $\hat{a}$ et $\hat{b}$ qui permettent de construire des surfaces projetées en utilisant simplement $\hat{k}_t = \hat{a}t + \hat{b}$. C'est ce modèle qui sera utilisé par la suite.

Il est précisé que l'objectif du présent travail étant de quantifier dans le risque total auquel est exposé un régime rentier la part issue de l'incertitude sur la mortalité future, nous privilégions le choix d'un modèle stochastique simple et opérationnel. Les limites du modèle de Lee-Carter si elles peuvent en effet avoir un impact sur le niveau absolu de la provision obtenue doivent avoir un impact moindre sur la répartition entre la part mutualisable et la part non mutualisable du risque de mortalité. Ce point est discuté en conclusion à ce travail.

On notera également que, compte tenu de la répartition actuelle par âge du groupe étudié, la méthode de fermeture de la table n'est pas déterminante dans la formation du résultat. Plus généralement on peut considérer que l'importance du choix d'une méthode de fermeture peut être relativisée dans le cas d'un régime de rentes. On pourra consulter sur ce point PLANCHET et THEROND [2006]. La question de la fermeture des tables est également discutée dans DENUIT et QUASHIE [2005] avec un point de vue sensiblement différent.

## 2.2. APPLICATIONS NUMERIQUES

On présente ici les résultats obtenus tout d'abord sur la famille de tables prospectives proposée puis, dans un second temps, les conséquences en terme de valorisation de l'engagement du régime de rentes.

### 2.2.1. Tables prospectives

La table prospective utilisée dans cette étude est construite à partir des tables du moment fournies par l'INED[2] dans MESLE et VALLIN [2002] ; le lien entre les taux de décès discrets et la version continue (sous l'hypothèse d'un taux de hasard constant sur chaque carré du diagramme de Lexis) nous permet d'obtenir les taux de hasard à ajuster à partir des données de base :

$$\hat{\mu}_{xt} = -\mathbf{ln}\left(1-\hat{q}_{xt}\right). \qquad (10)$$

L'ajustement sur les données historiques conduit à la surface de mortalité suivante :

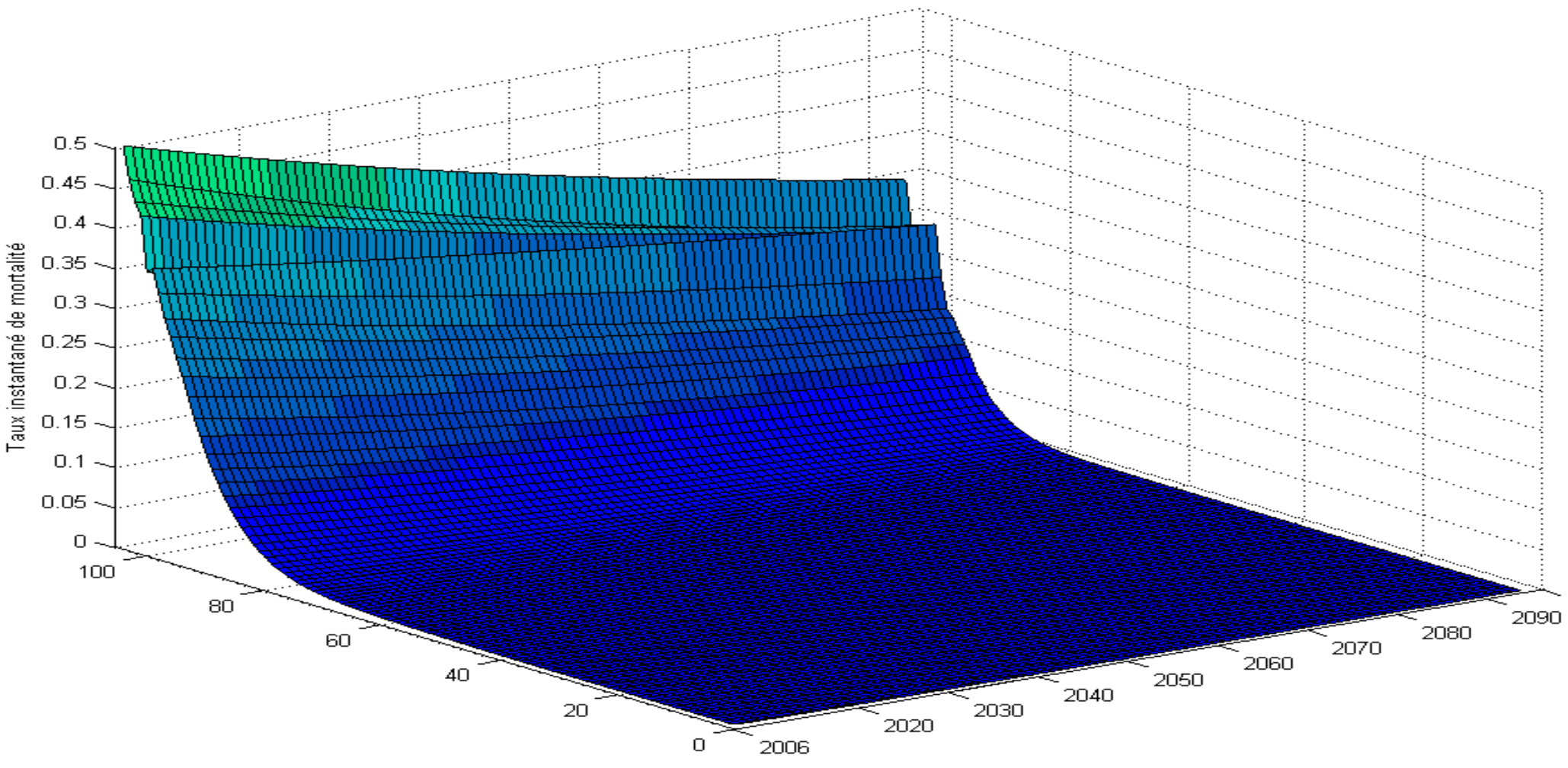

Fig. 4 : *Surface de mortalité ajustée*

En ce qui concerne le volet prédictif du modèle, nous obtenons sur nos données $\hat{b} = 4059{,}94439$, $\hat{a} = -2{,}05775$ et $\hat{\sigma}_{\gamma} = 3{,}9388782$.

[2] Ces tables sont disponibles sur http://www.ined.fr/publications/cdrom_vallin_mesle/Tables-de-mortalite/Tables-du-moment/Tables-du-moment-XX.htm

### 2.2.2. Engagement du régime

Au-delà des moments d'ordre un et deux de la distribution de $\Lambda$, qui peuvent être obtenus explicitement (*cf.* MAGNIN et PLANCHET [2000]), on s'intéresse ici à la distribution de l'engagement du régime. La méthode retenue consiste à simuler les durées de survie des rentiers, $T_{x(j)}, j \in J$ puis à calculer :

$$\Lambda = \sum_{t=1}^{\infty} \frac{1}{(1+i)^t} \sum_{j \in \boldsymbol{J}} r_j * \boldsymbol{1}_{]t;\infty[}\left(T_{x(j)}\right). \tag{11}$$

Sur la base des réalisations des $T_{x(j)}, j \in J$ obtenues. On obtient ainsi des réalisations $\lambda_1, ..., \lambda_n$ de $\Lambda$ et on détermine la fonction de répartition empirique de l'engagement, représentée ci-dessous (dans le cas de 20 000 simulations) :

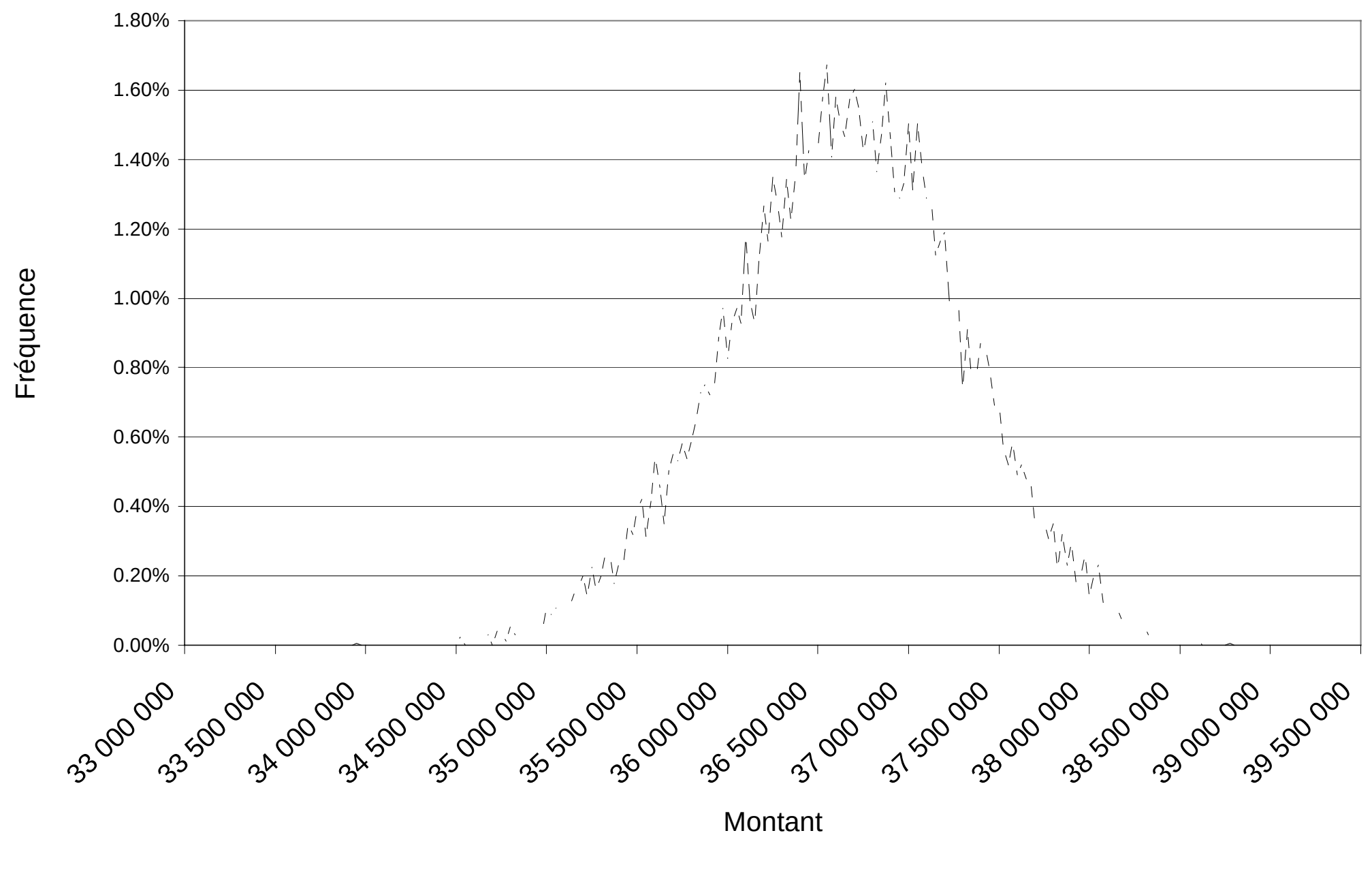

Fig. 5 : *Distribution empirique de l'engagement*

La provision $L_0$ est approchée par $\overline{\lambda} = \frac{1}{N} \sum_{n=1}^{N} \lambda_n$. La variance de l'engagement est estimée par $\frac{1}{N-1} \sum_{n=1}^{N} \left(\lambda_n - \mathrm{L}_0\right)^2$ ; on s'intéresse plus particulièrement au coefficient de variation empirique :

$$cv = \frac{\sqrt{\frac{1}{N-1}\sum_{n=1}^{N}\left(\lambda_n - \mathrm{L}_0\right)^2}}{\frac{1}{N}\sum_{n=1}^{N}\lambda_n}, \tag{12}$$

qui fournit un indicateur de la dispersion de l'engagement et dans une certaine mesure de sa « dangerosité ». On obtient pour ces différents indicateurs les valeurs suivantes :

| | **Déterministe** |
|---|---|
| Espérance | 36 653 830 |
| Ecart-type | 653 218 |
| Borne inférieure de l'intervalle de confiance | 33 957 502 |
| Borne supérieure de l'intervalle de confiance | 38 905 879 |
| Coefficient de variation | 1,78% |

On note malgré la taille modeste de la population que l'engagement est évalué avec une précision relative de plus ou moins 6,1 % (rapport de la demi-largeur de l'intervalle de confiance à l'espérance). Le risque démographique est de ce point de vue assez bien maîtrisé.

<u>Remarque</u> : Il est important de simuler les durées de survie de manière efficace ; on utilise donc ici la « méthode d'inversion » dans le contexte discret, qui repose sur le constat que si on définit la variable *T* par :

- ✓ $T=0$ si $U < p_0$
- ✓ T=1 si $p_0 \leq U < p_0 + p_1$
- ✓ ….
- ✓ T=$j$ si $\sum_{i=0}^{j-1} p_i \leq U < \sum_{i=0}^{j} p_i$

avec U une variable de loi uniforme sur $\left[0,1\right]$, alors *T* est distribuée selon la loi $(p_n)_{n\geq 0}$. Le lien entre les probabilités $p_i$ et les taux de décès déterminés précédemment est donné par $p_i = q_{x+i} \times \prod_{j=0}^{i-1}\left(1 - q_{x+j}\right)$. Par rapport à l'approche directe consistant à décider de la survie de chaque rentier par un tirage à chaque période, en comparant le résultat obtenu au taux de décès à l'âge considéré, on divise le temps de simulation par un facteur de l'ordre de 20. Cette optimisation est indispensable pour conserver un caractère opérationnel au modèle dans le contexte de la mortalité stochastique, examiné *infra*.

## 3. CAS D'UNE MORTALITE STOCHASTIQUE

On intègre ici au modèle une composante de risque systématique au travers d'une incertitude sur la surface de mortalité future, la surface construite à la section précédente définissant alors la tendance de référence autour de laquelle la mortalité observée sera supposée fluctuer.

### 3.1. MODELE DE MORTALITE

On utilise l'équation de régression qui a permis d'obtenir la tendance de mortalité future :

$$k_t^* = at + b + \gamma_t , \tag{13}$$

et l'on obtient des réalisations de la mortalité future en effectuant des tirages dans la loi du résidu $(\gamma_t)$, $N\left(0,\sigma_\gamma^2\right)$. La variable $k_t^*$ est donc gaussienne telle que $E\left(k_t^*\right) = k_t$. On obtient ainsi des réalisations des taux instantanés de sortie *via* :

$$\mu_{xt}^* = \mathbf{exp}\left(\alpha_x + \beta_x k_t^*\right). \tag{14}$$

Puisque $k_t^*$ est gaussienne, $\mu_{xt}^*$ est log-normale et sont espérance se calcule selon $E\left(\mu_{xt}^*\right) = E\,\mathbf{exp}\left(\alpha_x + \beta_x k_t^*\right) = \mathbf{exp}\left(\alpha_x + \beta_x k_t + \frac{\beta_x^2 \sigma_\gamma^2}{2}\right)$ et donc :

$$E\left(\mu_{xt}^*\right) = \mu_{xt}\,\mathbf{exp}\left(\frac{\beta_x^2 \sigma_\gamma^2}{2}\right) > \mu_{xt} . \tag{15}$$

Le modèle stochastique a donc tendance à surestimer les taux de sortie. Mais, comme l'illustre la figure 6, en pratique ce biais est faible et peu pénalisant pour le modèle, car peu significatif aux âges élevés. On représente sur la figure 6 $E\left(\mu_{xt}^*\right) / \mu_{xt}$ :

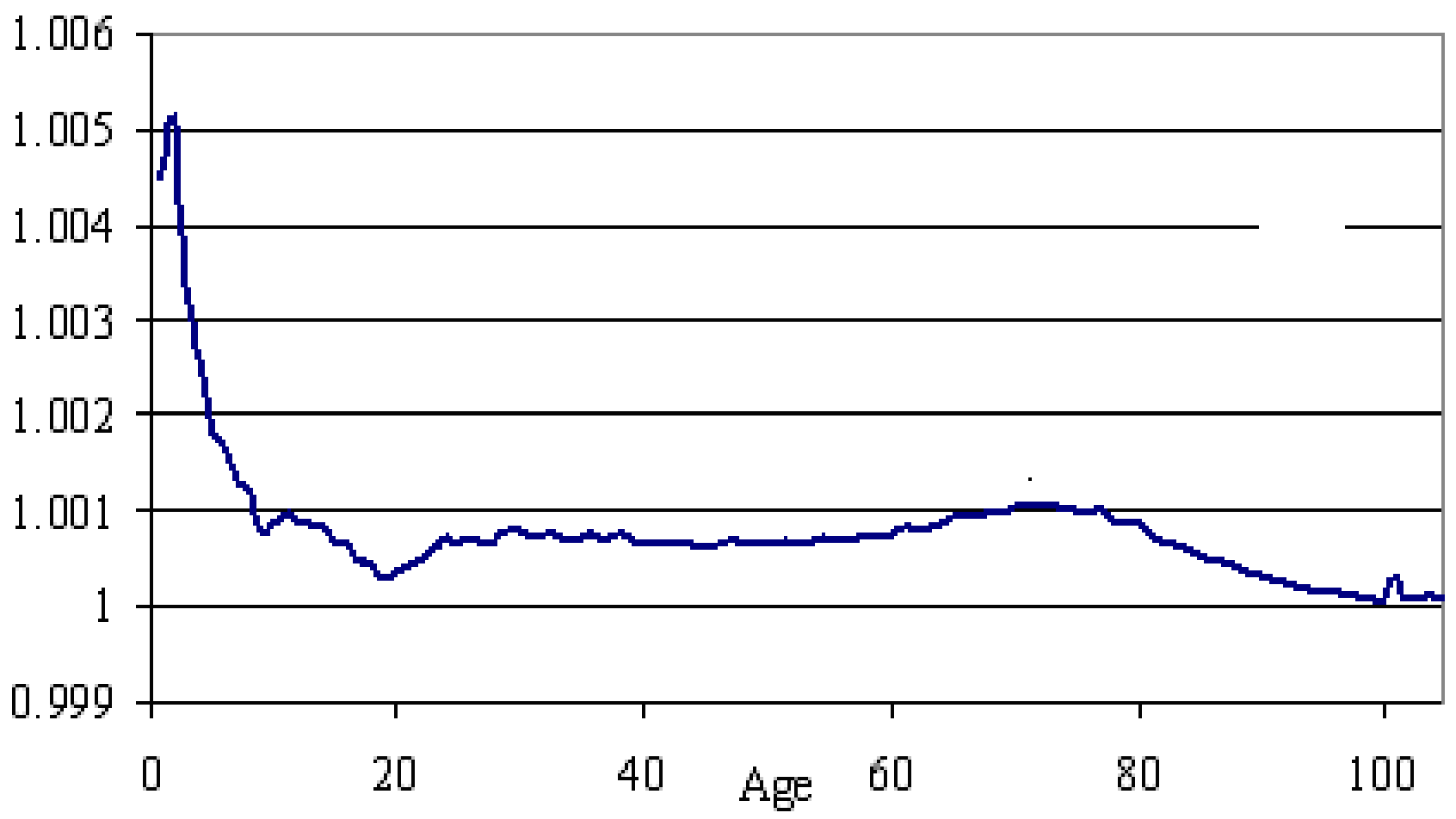

Fig. 6 : *Analyse du biais relatif en fonction de l'âge*

Incidemment, on on ne peut montrer simplement si le biais qui en résulte sur les taux discrets $q_x$ est positif ou négatif. En effet, l'inégalité de Jensen ($f(EX) \leq Ef(X)$) appliquée à $f(x) = -\left(1-e^{-x}\right)$ conduit à $E\left(q_x^*\right) \leq 1 - \mathbf{exp}\left(-E\mu_x^*\right)$, ce qui ne permet pas de conclure simplement sur le sens du biais. Cela indique toutefois que la relation entre $q_x$ et $\mu_x$ n'amplifie pas le biais, et donc le biais sur le taux discret doit être également faible.

On observera que la volatilité $\sigma_\gamma$ est un paramètre permettant de contrôler simplement le degré d'incertitude attaché à la mortalité future. Dans le cas particulier $\sigma_\gamma = 0$ on retrouve le modèle déterministe analysé à la section précédente.

Le biais généré sur les taux de sortie par le mode de simulation des surfaces de mortalité, s'il s'avère peu pénalisant au voisinage de la volatilité $\sigma_\gamma$ estimée sur les données disponibles, pourrait être potentiellement gênant dans des situations de plus forte volatilité : ce point sera illustré par une application numérique. Aussi, nous utiliserons dans la suite une version corrigée du biais du modèle proposé définie par :

$$\mu_{xt}^* = \mathbf{exp}\left(\alpha_x - \frac{\beta_x^2 \sigma_\gamma^2}{2} + \beta_x k_t^*\right). \tag{16}$$

Cette version du modèle satisfait par construction $E\left(\mu_{xt}^*\right) = \mu_{xt}$. Elle apparaît donc plus cohérente avec l'objectif recherché de « perturber » la surface de mortalité, mais sous l'hypothèse que celle-ci définit correctement la tendance future des taux instantanés de décès.

## 3.2. ANALYSE DE LA REPARTITION DU RISQUE POUR LE REGIME DE RENTES

On retient comme mesure du risque la variance de la somme des flux futurs actualisés $\Lambda$. En conditionnant par la surface de mortalité $\Pi$ et en utilisant l'équation de décomposition de la variance on obtient :

$$\boldsymbol{V}[\Lambda]=\boldsymbol{E}\left[\boldsymbol{V}(\Lambda|\Pi)\right]+\boldsymbol{V}\left[\boldsymbol{E}(\Lambda|\Pi)\right]. \tag{17}$$

Le second terme du membre de droite de l'expression ci-dessus représente le risque systématique associé au régime de rentes ; le premier terme le risque technique, *i.e.* le risque de mortalité mutualisable. En pratique on retiendra comme indicateur la part de variance expliquée par la composante de risque systématique, soit :

$$\omega(\sigma_\gamma)=\frac{\boldsymbol{V}\left[\boldsymbol{E}(\Lambda|\Pi)\right]}{\boldsymbol{V}[\Lambda]}. \tag{18}$$

Lorsque la taille du groupe tend vers l'infini, $\omega(\sigma_\gamma)$ converge vers 1. En d'autres termes pour un groupe parfaitement mutualisé, toute la variance est expliquée par la composante systématique. Le calcul direct de $\omega(\sigma_\gamma)$ n'est pas aisé, aussi on utilise ici une approche par simulation : on sélectionne dans un premier temps, au moyen d'un tirage dans une loi $N\left(0,\sigma_\gamma^2\right)$ une surface de mortalité (en pratique les réalisations de la gaussienne sont obtenues par inversion de la fonction de répartition en utilisant l'approximation de Moro présentée dans PLANCHET et al. [2005]) ; puis, conditionnellement à cette hypothèse de mortalité, on simule la survie des rentiers (« le passif »).

De manière plus formelle, si $\lambda_{n,m}$ est la réalisation de $\Lambda$ résultant de la $n$-ième trajectoire de la mortalité et de la $m$-ième trajectoire du passif, notons :

$$\overline{\lambda}_n=\frac{1}{M}\sum_{m=1}^{M}\lambda_{n,m} \text{ et } \overline{\overline{\lambda}}=\frac{1}{N}\sum_{n=1}^{N}\overline{\lambda}_n=\frac{1}{NM}\sum_{n=1}^{N}\sum_{m=1}^{M}\lambda_{n,m}. \tag{19}$$

Les quantités ci-après sont des estimateurs sans biais et convergents respectivement de $\boldsymbol{E}\left[\boldsymbol{V}(\Lambda|M)\right]$ et de $\boldsymbol{V}\left[\boldsymbol{E}(\Lambda|M)\right]$ :

$$\hat{\boldsymbol{E}}\left[\boldsymbol{V}\left(\Lambda|M\right)\right]=\frac{1}{N}\sum_{n=1}^{N}\frac{1}{M-1}\sum_{m=1}^{M}\left(\lambda_{n,m}-\overline{\lambda}_n\right)^2, \tag{20}$$

$$\hat{\boldsymbol{V}}\left[\boldsymbol{E}\left(\Lambda|M\right)\right]=\frac{1}{N-1}\sum_{n=1}^{N}\left(\overline{\lambda}_n-\overline{\overline{\lambda}}\right)^2. \tag{21}$$

Le calibrage du nombre de simulations est effectué de manière empirique en arrêtant l'algorithme lorsque les résultats se stabilisent (écart entre deux résultats successifs $<10^{-3}$).

### 3.3. APPLICATION NUMERIQUE

Les applications numériques proposées sont à deux niveaux : dans un premier temps on analyse l'impact de la prise en compte de la mortalité stochastique sur la loi de l'engagement du régime considéré ; puis, dans un second temps, on détermine l'évolution en fonction de la volatilité de la surface de mortalité (et donc de l'incertitude attachée à la prévision de cette dernière) de la part de variance expliquée par cette source de risque dans la variance totale.

#### 3.3.1. Analyse de l'engagement

On détermine tout d'abord pour la valeur estimée de la volatilité de la surface de mortalité ($\sigma_\gamma=3{,}94$) la distribution empirique de l'engagement, représentée ici avec la distribution de référence obtenue précédemment dans le cas déterministe (avec 20 000 tirages) :

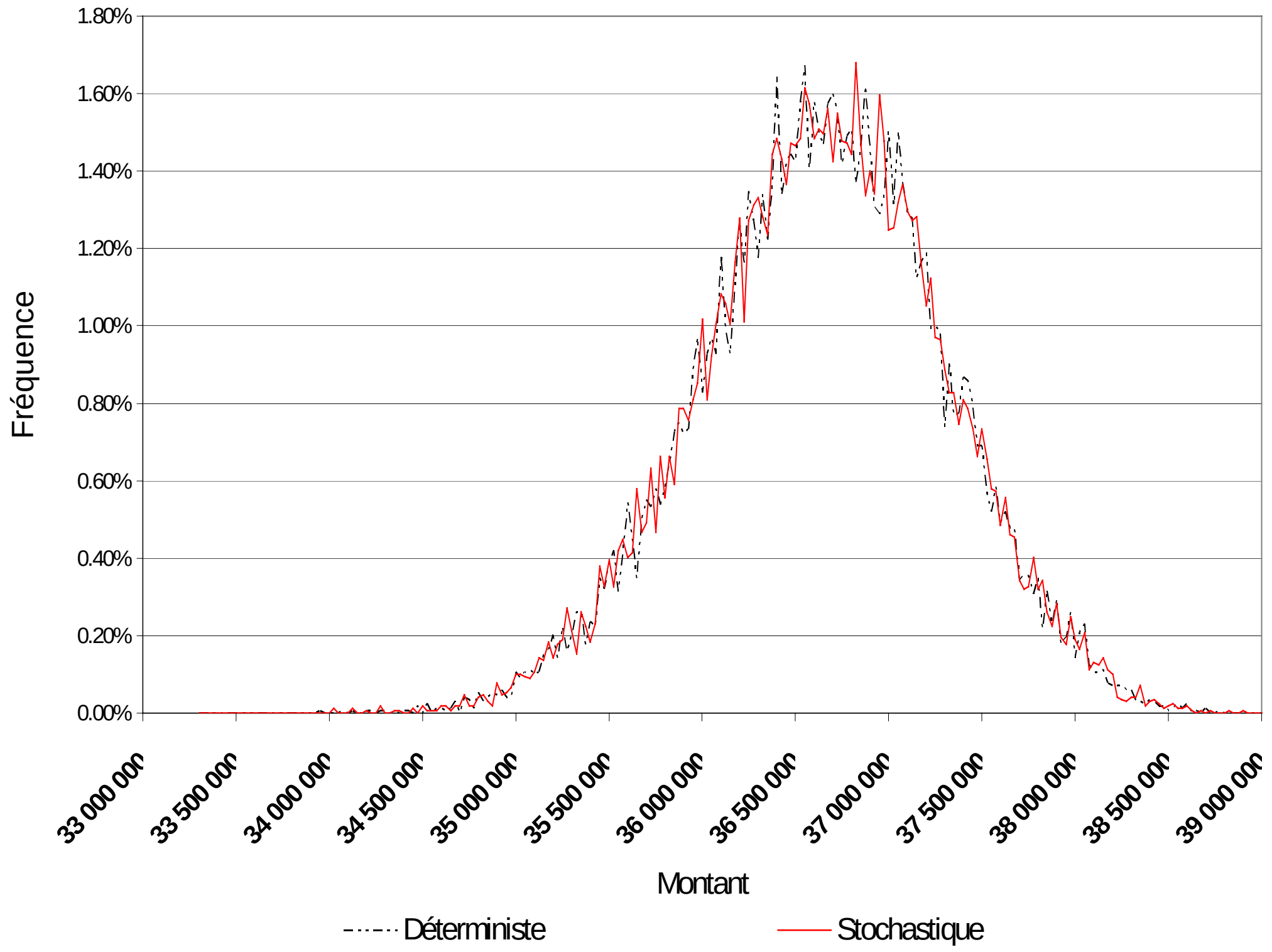

Fig. 7 : *Distribution empirique de l'engagement : approche déterministe VS approche stochastique*

On peut noter que le coefficient de variation de la distribution de l'engagement est estimé à 1,79 %, dans les deux situations (mortalité déterministe et mortalité stochastique) : alors que l'on aurait pu s'attendre à ce que la prise en compte de ce facteur de risque supplémentaire augmente sensiblement la dispersion, et donc la « dangerosité » de l'engagement, on constate ici une quasi-stabilité. Les résultats détaillés sont repris ci-après :

| | **Déterministe** | **Stochastique** |
|---|---|---|
| Espérance | 36 653 830 | 36 654 550 |
| Ecart-type | 653 218 | 655 180 |
| Borne inférieure de l'intervalle de confiance | 33 957 502 | 34 028 507 |
| Borne supérieure de l'intervalle de confiance | 38 905 879 | 38 924 564 |
| Coefficient de variation | 1,78% | 1,78% |

On note en particulier que l'espérance de l'engagement est identique dans le cas déterministe et dans le cas stochastique, ce qui illustre le faible biais induit par le mode de sélection des tables aléatoires au travers de $k_t$. Dans le cas présent, les données de référence utilisées pour calibrer la mortalité de référence fluctuent faiblement, ce qui conduit à un risque associé marginal au regard du risque d'échantillonnage. Ceci est d'autant plus marqué que la population concernée est d'effectif relativement faible.

Dans une approche « valeur à risque » (*VaR*), on trouve que le quantile à 75 % de la distribution de l'engagement est de 37,1 M€ dans le cas stochastique, ce qui est quasiment identique à la valeur obtenue dans le cas déterministe.

Ces résultats sont bien entendu la conséquence directe de la très faible volatilité estimée sur la série $k_t$ ; dans le cas où cette volatilité augmente, la distribution de l'engagement intégrant le risque systématique diverge rapidement de la situation de référence constituée par le modèle déterministe. A titre d'illustration, on présente ci-après les résultats avec une volatilité multipliée par 10 (le nombre de tirages reste fixé à 20 000) :

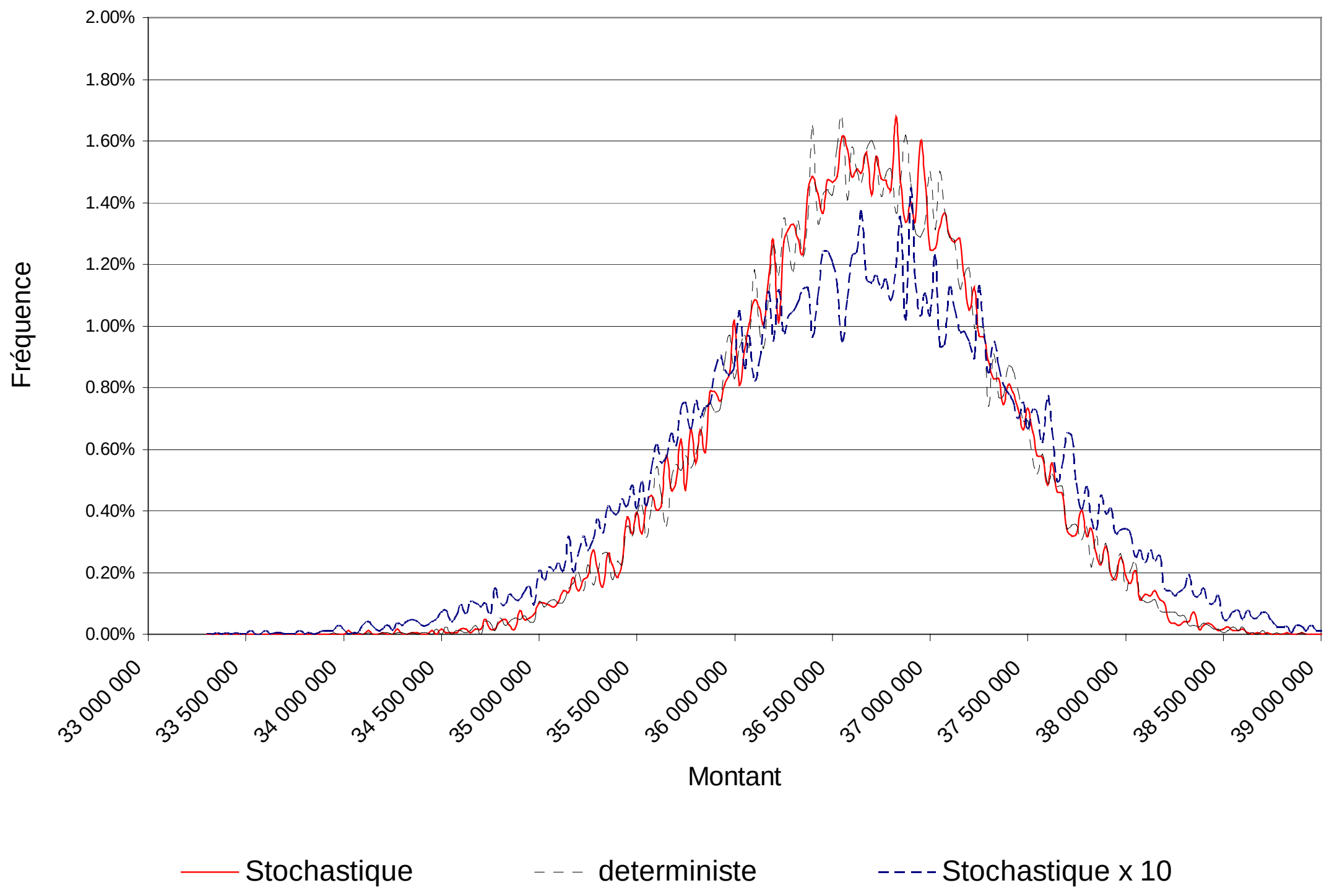

Fig. 8 : *Distribution empirique de l'engagement : approche déterministe VS approche stochastique*

On peut remarquer que le coefficient de variation de la distribution de l'engagement dans le cas de la prise en compte d'un caractère stochastique de la mortalité à forte volatilité est estimé à 2,18 %, contre 1,79 % dans la situation de référence : le modèle, comme on pouvait s'y attendre, est très sensible à la volatilité de processus de génération des surfaces de mortalité stochastiques. Les résultats détaillés sont repris ci-après :

| | **Déterministe** | **Stochastique** | **Stochastique x10** |
|---|---|---|---|
| Espérance | 36 653 830 | 36 654 550 | 36 639 628 |
| Ecart-type | 653 218 | 655 180 | 833 047 |
| Borne inférieure de l'intervalle de confiance | 33 957 502 | 34 028 507 | 33 350 742 |
| Borne supérieure de l'intervalle de confiance | 38 905 879 | 38 924 564 | 39 457 435 |
| Coefficient de variation | 1,78% | 1,78% | 2,27% |

Bien entendu, toutes choses égales par ailleurs, l'impact sur la loi de l'engagement de la prise en compte de la composante non mutualisable du risque est d'autant plus important que les fluctuations d'échantillonnage sont faibles, et donc que le groupe est de taille importante. Afin d'illustrer ce point, les calculs de la distribution de l'engagement ont été menés sur un groupe fictif composé de 100 répliques du groupe de base, ce qui constitue ainsi un groupe de 37 400 rentiers. On obtient le graphique suivant :

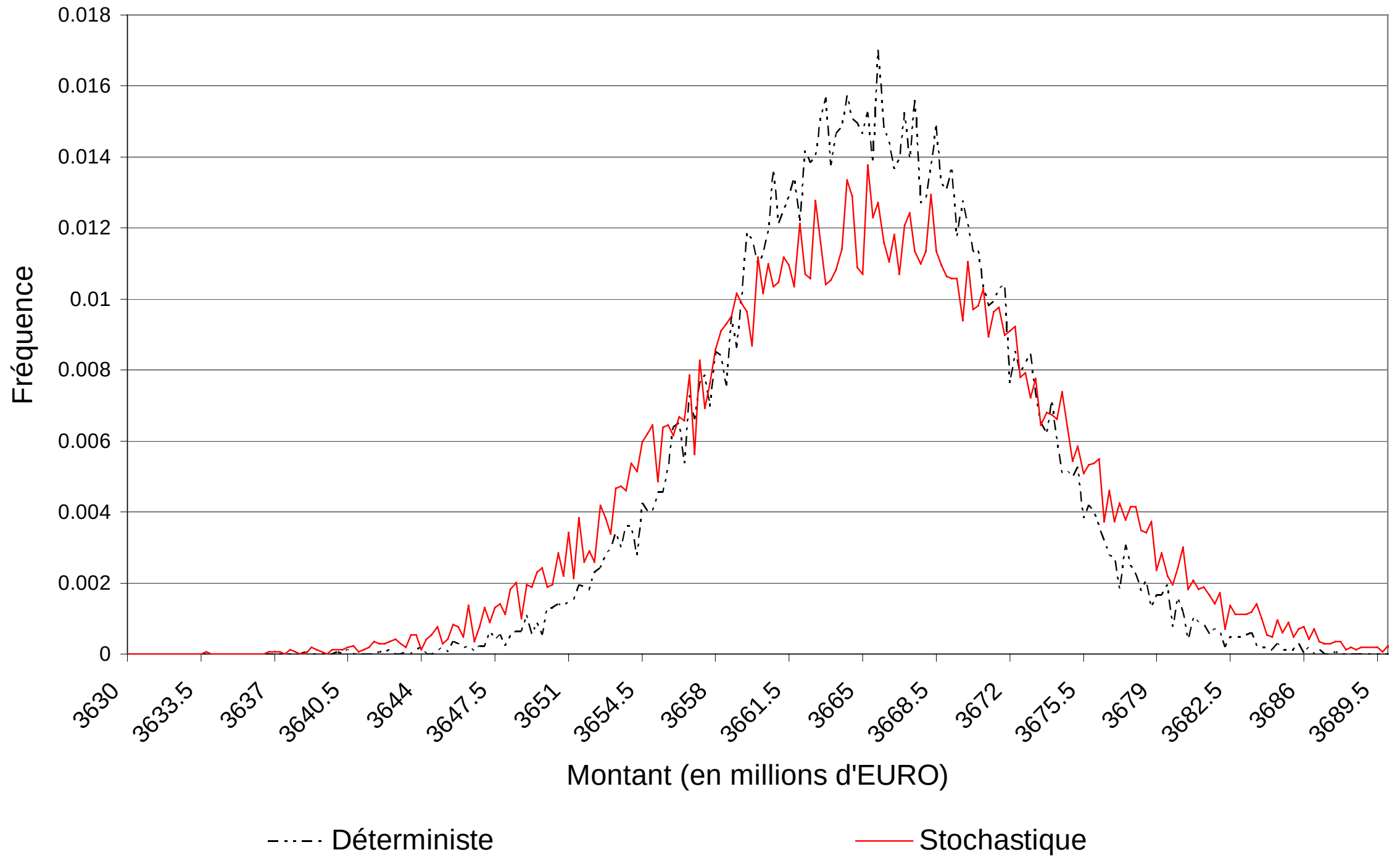

Fig. 9 : *Distribution empirique de l'engagement : taille du groupe x100*

On constate sur le graphique l'applatissement de la distribution de l'engagement, illustrant une situation *a priori* plus risquée pour le régime lorsque la part non mutualisable du risque est prise en compte. Les résultats synthétiques sont détaillés ci-dessous :

| | **Déterministe** | **Stochastique** |
|---|---|---|
| Espérance | 3 665 313 406 | 3 665 305 670 |
| Ecart-type | 6 547 525 | 8 224 853 |
| Quantile à 5% | 3 654 250 000 | 3 652 000 000 |
| Quantile à 95% | 3 675 750 000 | 3 678 750 000 |
| Coefficient de variation | 0,179% | 0,224% |

On constate que si l'espérance de l'engagement est bien simplement multipliée par 100 par rapport à la situation de référence, on voit apparaître une différence sensible entre l'approche déterministe classique et l'approche stochastique : l'écart-type augmente en effet de 25 % lorsque l'on prend en compte le risque non mutualisable.

On peut toutefois noter que le risque démographique, même en présence du facteur de risque systématique, reste bien maîtrisé, l'intervalle de confiance à 95 % pour le niveau de l'engagement étant (en millions d'euros) $\left[3\,637;3\,681\right]$, ce qui donne une précision dans la mesure de l'engagement supérieure à 99 %.

Dans une approche « valeur à risque » (*VaR*), on trouve que le quantile à 75 % de la distribution de l'engagement est de 3 671 M€ dans le cas stochastique, contre 3 669 M€ dans le cas déterministe : là encore on note un très faible impact absolu sur le régime.

### 3.3.2. Importance de la correction de biais

Dès lors que la volatilité du processus de mortalité sous-jacent devient significative, la correction de biais s'avère indispensable. Le graphique suivant, qui représente la distribution empirique avec une volatilité égale à 10 fois, puis 20 fois la volatilité de référence du modèle sans correction du biais illustre ce point :

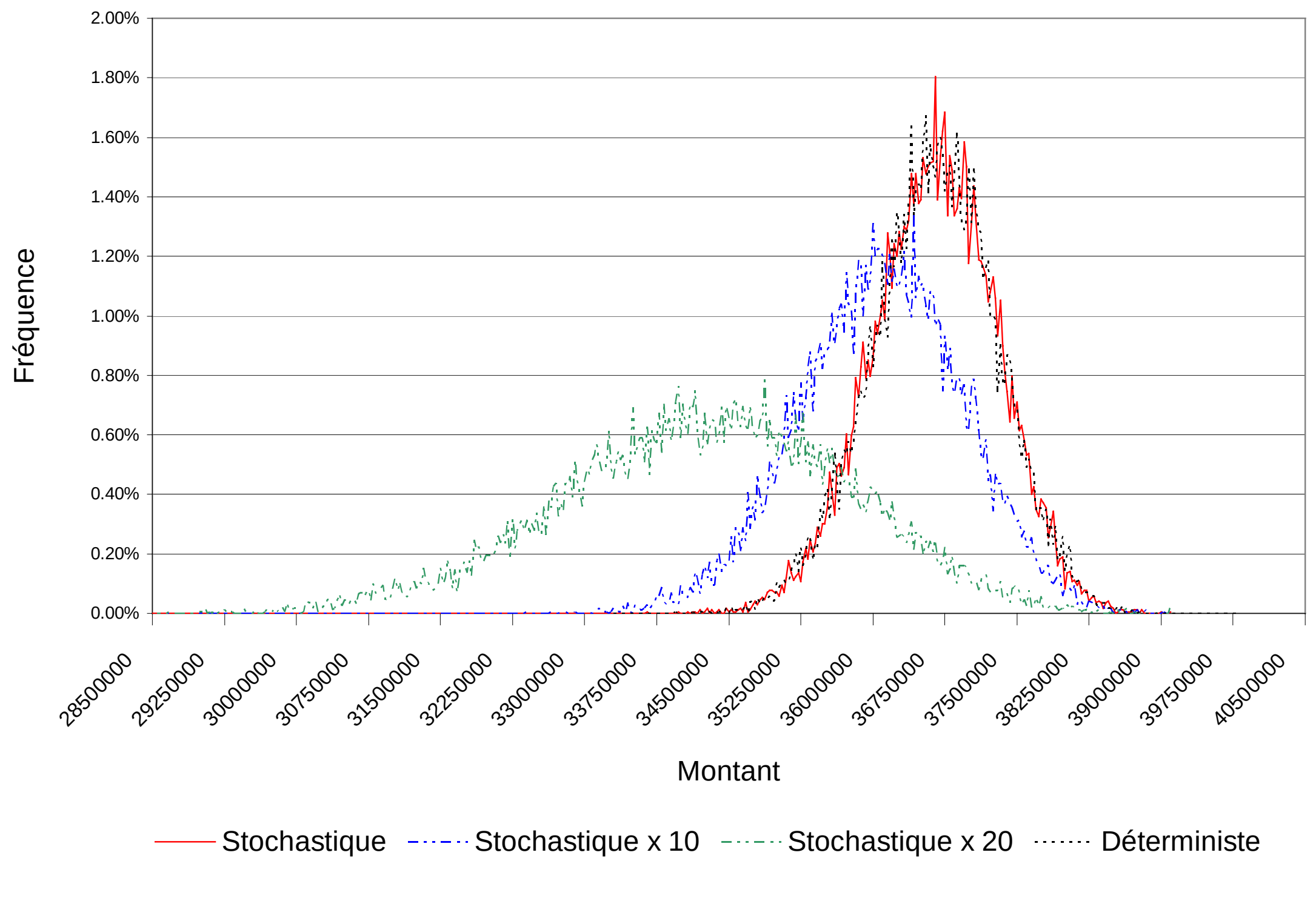

Fig. 10 : *Distribution empirique de l'engagement : approche déterministe VS approche stochastique*

On note que la distribution intégrant le risque systématique se trouve décalée vers la gauche du fait du biais sur les taux de sortie et qu'elle conduit ainsi à une vision minorée de l'engagement.

### 3.3.3. Analyse de la part de variance expliquée

Comme les résultats précédents l'ont montré, le comportement de l'engagement est fonction de la volatilité du processus stochastique de mortalité sous-jacent, qu'il importe donc de quantifier. Les estimateurs présentés *supra* nous permettent d'obtenir le graphique d'évolution de la part du risque systématique dans le risque global $\omega\left(\sigma_\gamma\right)$ en fonction de la volatilité $\sigma_\gamma$ du générateur de tables prospectives.

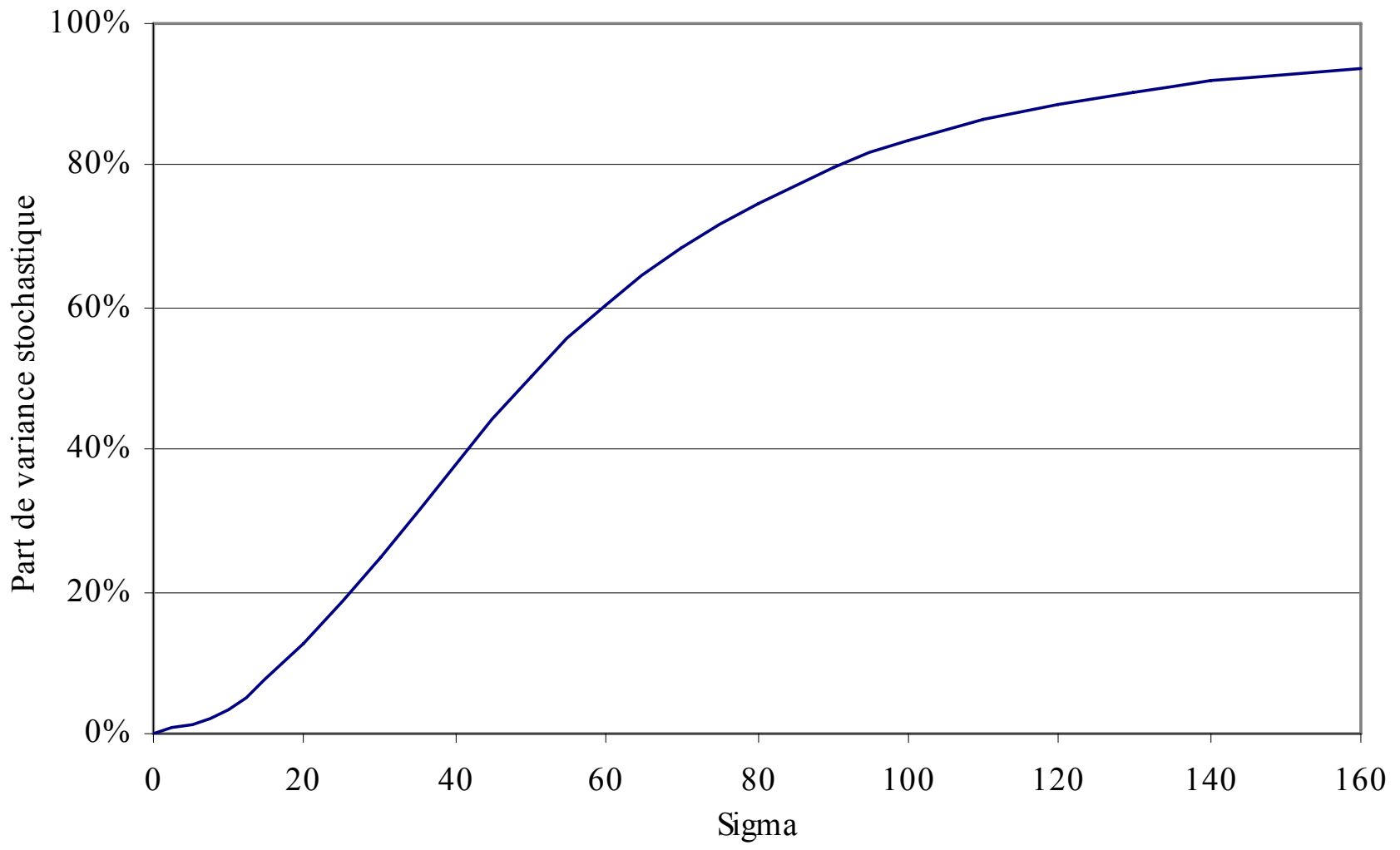

Fig. 11 : *Evolution de la part du risque systématique dans le risque global en fonction de $\sigma_\gamma$*

On note que la croissance de la courbe est lente au début et ne s'accélère que pour les valeurs de la volatilité importantes.

A $\sigma_\gamma$ fixé, la part de variance expliquée par la composante stochastique de la mortalité augmente avec la taille du portefeuille ; on obtient ainsi (la taille du portefeuille est exprimée en nombre de fois la taille du portefeuille de référence) :

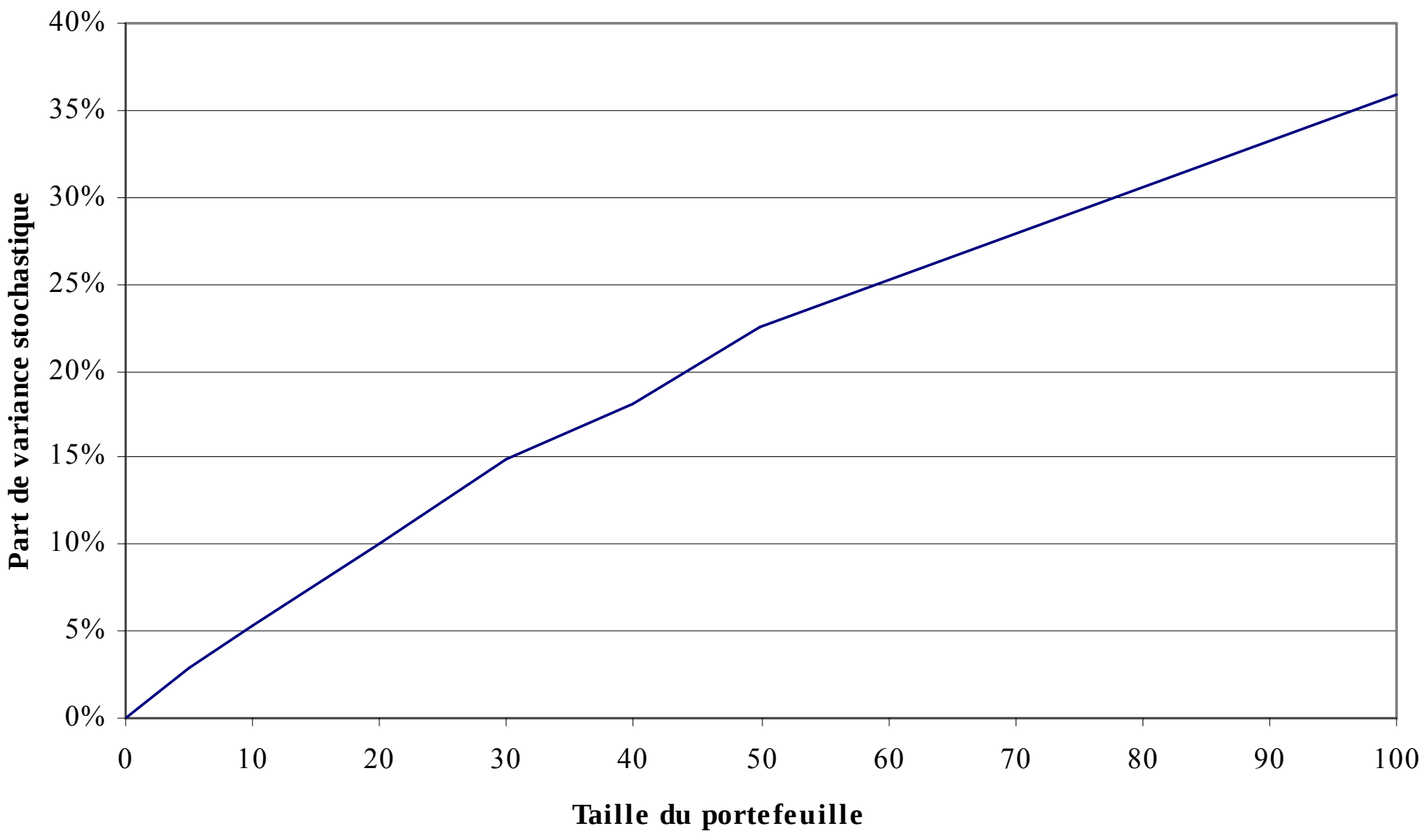

Fig. 12 : *Evolution de la part du risque systématique dans le risque global en fonction de la taille du portefeuille*

On note une croissance relativement lente (linéaire) de cette source de risque dans la variance totale, le seuil d'un tiers étant par exemple atteint pour un régime de plus de 300 000 rentiers ; pour des régimes de cette taille les fluctuations d'échantillonnage sont très faibles, et la variance globale est donc petite.

### 3.4. MESURE DU RISQUE D'ERREUR DE DERIVE

Le régime considéré ici est en fait soumis à deux risques distincts : le premier, dont l'étude occupe le présent article, est la conséquence des fluctuations aléatoires des taux de mortalité futurs autour de la tendance définie par la table prospective. Le second est quant à lui associé à l'incertitude sur cette tendance.

Plus précisément, cette incertitude a elle-même deux sources : l'erreur de modèle et l'imprécision associée à la détermination des coefficients de l'interpolation.

Il est délicat d'évaluer *a priori* l'erreur de modèle ; une première approche consiste par exemple à tester la robustesse du modèle proposé en effectuant les estimations sur différentes plages temporelles. Ces estimations conduisent à des estimations de la mortalité future qui peuvent être sensiblement divergentes.

L'imprécision attachée aux coefficients de l'interpolation eux-mêmes est plus aisée à quantifier. En effet, dans le modèle :

$$k_t^* = at + b + \gamma_t, \quad (22)$$

les coefficients $a$ et $b$ sont estimés par la méthode des moindres carrés ordinaires, qui conduit à ce que le couple $\left(\hat{a},\hat{b}\right)$ soit un vecteur gaussien. On peut donc simuler différentes droites d'extrapolation et intrégrer cette source de variabilité dans le modèle.

Ce développement dépasse le cadre du présent article et fera l'objet d'un travail spécifique ultérieur. Il est toutefois important d'avoir à l'esprit l'existence de cette seconde source de risque, potentiellement beaucoup plus importante pour le régime, les écarts de pente sur la droite d'extrapolation conduisant à des taux de survie de plus en plus divergents au cours du temps.

## 4. CONCLUSION

On a mis en œuvre dans le présent travail un modèle opérationnel pour quantifier la part de risque systématique de mortalité dans l'engagement d'un régime de rentes. Les résultats obtenus tendent à montrer qu'en-deça d'un effectif de quelques dizaines de milliers de têtes ce risque est négligeable en comparaison des fluctuations d'échantillonnage, ces dernières expliquant alors l'essentiel de la variance.

Pour les gros portefeuilles, si le risque systématique devient significatif d'un point de vue relatif, le risque mutualisable diminuant avec l'effectif concerné, le niveau absolu de risque reste faible. Au global la largeur de l'intervalle de confiance à 95% pour l'estimation de l'engagement est décroissante avec la taille de la population sous risque, même en intrégrant le risque systématique.

La principale limite du modèle proposé est la conséquence directe d'une limitation de la modélisation de Lee-Carter (ou des modèles poissoniens) : les taux de mortalité aux différents âges sont en effet supposés parfaitement corrélés, la composante aléatoire $(k_t)$ ne dépendant que du temps[3], et le taux de sortie instantané étant obtenu par la formule $\ln \mu_{xt} = \alpha_x + \beta_x k_t$. Ceci est en pratique contredit par les observations (*cf.* figure 1).

Toutefois l'approche proposée dans le présent travail peut être aisément adaptée à l'ensemble des modèles reposant sur une modélisation de la tendance temporelle au travers d'une série temporelle (éventuellement sur plusieurs paramètres). Une présentation d'un certain nombre d'autres modèles de ce type (et notamment les approches utilisant les logits des taux de décès) est effectuée dans PLANCHET [2005] et dans SERANT [2005].

[3] Cette composante est modélisée par un processus ARIMA.

Il nous semble utile de préciser que ce type d'approche fournit un cadre opérationnel pour répondre aux exigences des futures dispositions Solvabilité 2 (et également dans le contexte de l'IFRS 4 phase 2, quoi que sur ce point les normes comptables soient moins exigeantes).

On soulignera toutefois que, si ces approches permettent de quantifier le risque associé au caractère intrinsèquement aléatoire de la mortalité future, autour d'une tendance supposée prévisible, elles n'apportent pas pour autant de réponse à une éventuelle erreur de spécification sur la tendance de référence. On constate ainsi par exemple que les modèles prospectifs construits par le passé ont systématiquement sous-estimé l'augmentation de l'espérance de vie des rentiers, et donc le niveau des provisions d'un régime de rentes. Des modèles strictements extrapolatifs peuvent donc apparaître limités de ce point de vue.

On peut envisager, face aux incertitudes sur l'évolution de la mortalité future, d'introduire dans les modèles une contrainte exogène permettant d'intégrer une opinion *a priori* que l'on peut avoir sur la mortalité future, indépendemment de la tendance passée. Une piste, en cours d'exploration et qui donnera lieu à un prochain article, consiste à imposer une contrainte sur l'espérance de vie à un âge de référence et à contrôler son évolution de manière exogène.

Une seconde piste en cours consiste à modéliser l'erreur dans l'appréciation de la dérive et de quantifier l'impact de cette erreur sur la distribution de l'engagement.

Enfin, à un moment où l'intérêt pour les problématiques de retraite s'accroit, ces modèles fournissent un outil supplémentaire aux régimes dans la mesure, et donc le pilotage, de leur équilibre technique.

## BIBLIOGRAPHIE

BROUHNS N., DENUIT M. [2001] « Risque de longévité et rentes viagères, Tables de mortalité prospectives pour la population belge », Discussion Paper, Institut de Statistique, Université catholique de Louvain, Louvain-la-Neuve, Belgique.

BROUHNS N., DENUIT M., VERMUNT J.K. [2002] « A Poisson log-bilinear regression approach to the construction of projected lifetables », *Insurance: Mathematics and Economics*, vol. 31, 373-393.

CAIRNS A., BLAKE D., DOWD K. [2004], « Pricing Frameworks for Securitization of Mortality Risk », *Proceedings of the 14$^{th}$ AFIR Colloquium*, vol. 1, 509-540.

CURRIE I.D., DURBAN M., EILERS P.H.C. [2004] « Smoothing and forecasting mortality rates », *Statistical Modelling*, vol. 4, 279-298.

DAHL M. [2004] « *Stochastic mortality in life insurance: market reserves and mortality-linked insurance contracts* ». Insurance: Mathematics and Economics, vol. 35, n°1, 113-136.

DEBON A., MARTINEZ-RUIZ F., MONTES F. [2004] « Dynamic Life Tables: A Geostatistical Approach », IME Congress.

DENUIT M., QUASHIE A. [2005] « Modèles d'extrapolation de la mortalité aux grands âges », *Institut des Sciences Actuarielles et Institut de Statistique Université Catholique de Louvain*, Louvain-la-Neuve, Belgique.

FAUCILLON L., JUILLARD M., LUONG TIEN D., LUU ANH T., VO TRAN H. [2006] « Etude du risque systématique de mortalité », Rapport de groupe de travail, ISFA.

GUTTERMAN S., VANDERHOOFT I.T. [1999] « Forecasting changes in mortality: a search for a law of causes and effects », *North American Actuarial Journal*, vol. 2, 135-138.

KIMELDORF G.S, JONES D.A. [1967] « *Bayesian graduation* », TSA, XIX.

LEE R.D., CARTER L. [1992] « Modelling and forecasting the time series of US mortality », *Journal of the American Statistical Association*, vol. 87, 659–671.

LEE R.D. [2000] « The Lee–Carter method of forecasting mortality, with various extensions and applications », *North American Actuarial Journal*, vol. 4, 80–93.

LELIEUR V. PLANCHET F. [2006] *Construction de tables de mortalité prospectives : le cas des petites populations*, à paraître dans le *Bulletin Français d'Actuariat*, vol. 6, n°13.

MAGNIN F., PLANCHET F. [2000] « L'engagement d'un régime de retraite supplémentaire à prestations définies », *Bulletin Français d'Actuariat*, Vol. **4,** n°7.

MESLE F., VALLIN J. [2002] « Comment améliorer la précision des tables de mortalité aux grands âges ? Le cas de la France », *Population n°4*, INED, 603.

PETAUTON P. [1996] *Théorie et pratique de l'assurance vie*, Paris : DUNOD.

PLANCHET F. [2005] « Tables de mortalité d'expérience pour des portefeuilles de rentiers », note méthodologique de l'Institut des Actuaires.

PLANCHET F., THEROND P.E. [2004] « Allocation d'actifs d'un régime de rentes en cours de service ». *Proceedings of the 14$^{th}$ AFIR Colloquium*, vol. 1, 111-134.

PLANCHET F., THÉROND P.E., JACQUEMIN J. [2005] *Modèles financiers en assurance – analyses de risques dynamiques*, Paris : Economica.

PLANCHET F., THÉROND P.E. [2006] *Modèles de durée – applications actuarielles*, Paris : Economica.

RENSHAW A.E. [1991] « Actuarial graduation practice and generalized linear and non-linear models », *Journal of the Institute of Actuaries*, vol. 118, 295–312.

SAPORTA G. [1990] *Probabilités, analyse des données et statistique*, Paris : Editions Technip.

SCHRAGER D.F. [2006] « *Affine Stochastic Mortality* », IME, vol. 38, n°1, 81-97.

SERANT D. [2005] « Construction de tables prospectives de mortalité », Document interne Fédération Française des Sociétés d'Assurance.

SITHOLE T., HABERMAN S., VERRALL R.J. [2000] « An investigation into parametric models for mortality projections, with applications to immediate annuitants and life office pensioners », *Insurance: Mathematics and Economics*, vol. 27, 285–312.

TAYLOR G. [1992] « A bayesian interpretation of Whittaker-Henderson graduation », *Insurance: Mathematics and Economics*, vol. 11, 7-16.